\documentclass[apj]{emulateapj}
\usepackage{amsmath}

\def\msun{M$_{\odot}$}

\def\feh{[Fe/H]}
\def\fe{[Fe/H]}

\def\teff{$T_{\rm eff}$}
\def\vsini{$V_{\rm rot}\sin{i}$}

\def\mv{$\mathcal{M}_{\rm V}$}
\def\info{$\mathcal{I}$}
\def\y2{Y$^2$}

\def\kep{\emph{Kepler}}

\begin{document}
\title{Retired A Stars: The Effect of Stellar Evolution on the Mass
  Estimates of Subgiants}   

\author{
John Asher Johnson\altaffilmark{1,2},
Timothy D. Morton\altaffilmark{1,2},
Jason T. Wright\altaffilmark{3,4}
}

\email{johnjohn@astro.caltech.edu}

\altaffiltext{1}{Department of Astrophysics,
  California Institute of Technology, MC 249-17, Pasadena, CA 91125}
\altaffiltext{2}{NASA Exoplanet Science Institute (NExScI), CIT Mail
  Code 100-22, 770 South Wilson Avenue, Pasadena, CA 91125}
\altaffiltext{3}{Department of Astronomy and Astrophysics, 525 Davey Lab, The 
  Pennsylvania State University, University Park, PA 16803}
\altaffiltext{4}{Center for
  Exoplanets and Habitable Worlds, 525 Davey Lab, The 
  Pennsylvania State University, University Park, PA 16803} 

\begin{abstract}
Doppler surveys have shown that the occurrence rate of Jupiter-mass
planets appears to increase as a function of stellar mass.  However,
this result depends on the ability to accurately measure
the masses of evolved stars.  Recently, Lloyd (2011) called into
question the masses of subgiant stars targeted by Doppler surveys. Lloyd
argues that very few observable subgiants have masses greater than
1.5\msun, and that most of them have masses in the range
1.0-1.2~\msun. To investigate this claim, we use Galactic stellar
population models to generate an all-sky distribution of stars. We
incorporate the effects that make massive subgiants less numerous,
such as the initial mass function and differences in stellar evolution
timescales. We find that these effects lead to negligibly small
systematic errors in stellar mass estimates, in contrast to the
$\approx50$\% 
errors predicted by Lloyd. Additionally, our simulated target sample
does in fact include a significant fraction of stars with masses
greater than 1.5~\msun, primarily because the inclusion of an apparent
magnitude limit results in a 
Malmquist-like bias toward more massive stars, in contrast to the
volume-limited simulations of Lloyd. The magnitude limit shifts
the mean of our simulated distribution toward higher masses and
results in a relatively smaller number of evolved stars with masses in
the range 1.0--1.2~\msun. We conclude that, within the context of our
present-day understanding of stellar structure and evolution, many of
the subgiants observed in Doppler surveys are indeed as massive as
main-sequence A stars.
\end{abstract}

\keywords{Stars: evolution---Stars: fundamental parameters---Stars: general---(Stars): planetary systems}

\section{Introduction}

Studies of the relationships between exoplanets and their host stars
provide valuable clues about how planets form, and also point the way
to new discoveries. For example, the well-established relationship
between the occurrence rate of gas giant planets and host-star
metallicity \citep{santos04,fischer05b,johnson09b} may be an
indication that the formation  timescale for close-in giant planets
($a < 5$~AU) is shortened by the metal-enhancement, and hence
dust-enhancement, of protoplanetary disks \citep[e.g.][]{ida04}. For
this reason, certain Doppler surveys have biased their target lists
toward metal-rich stars, which has resulted in the discovery of many
of the known hot Jupiter systems \citep{fischer05a, bouchy05, sato05}. 

More recent Doppler surveys have discovered that stellar mass is
another key predictor of giant planet occurrence
\citep{johnson07b,johnson10c}. This relationship is based on Doppler
surveys of M dwarfs on one side of the stellar mass range 
\citep[e.g.][]{johnson10a}, and the evolved   counterparts of F- and
A-type stars on the more massive end
\citep{johnson07,lovis07,sato07}. These so-called ``retired A-stars''
exhibit dramatically slower rotation velocities (\vsini)
than their main-sequence progenitors \citep{gray85,donascimento00},
making them better targets for Doppler-based planet surveys
compared to their F- and A-type main-sequence counterparts
\citep{hatzes03,fischer03,galland05}.  

However, the mass estimates of subgiants targeted by Doppler surveys
have recently been called into question by \citet[][hereafter
  L11]{lloyd11}. In an attempt to study the effects of star-planet
tidal interactions in planetary systems with evolved host stars, L11
investigated the expected mass distribution of evolved stars near the
subgiant branch. By using stellar evolution model
grids, assumptions about the 
metallicity distribution in the Galaxy, and the form of the stellar
initial mass function (IMF), L11 concluded that most bright subgiants
are not the evolved brethren of A-type stars, but rather the evolved
counterparts of Sun-like stars. This is because massive stars evolve
much more quickly along the subgiant branch than do less massive
stars. As L11 notes, this differential evolution rate for stars of
different masses is a robust feature of stellar models. L11 predict
that this effect, together with the distribution of stellar masses
produced by the initial mass function, should  result in a very
small number of  massive 
subgiants with $M \gtrsim 1.5$~\msun\ in Doppler
surveys. We note that while L11 discuss stellar rotation in great
detail, it is this evolution rate feature that is his key argument
that subgiant masses are incorrect. We therefore focus our
investigation on this effect with the goal of assessing the question:
could the mass estimates of subgiants be systematically overestimated
by ignoring the stellar IMF and mass-dependent evolution rate along
the subgiant branch?

In this contribution we assess the specific critique of L11 using a
simple application of a Bayesian framework to the Galactic population
models of \citet{girardi05}.  We show that the neglect of the IMF and
the mass-dependent evolution timescales of subgiants results in a
small bias in the mass measurement towards higher 
masses. But that this bias is too small to cast doubt on the
conclusions of \citet{johnson10b, johnson10c}, namely that the
occurrence of Jovian planets increases with increasing stellar mass.
We also demonstrate that the mass 
distribution of the stars in the Johnson et al. Keck Doppler
survey is expected to contain a substantial number of subgiants,
consistent with the masses measured for that survey and strongly
inconsistent with the mass distribution predicted for it by L11. 

\section{Estimating the Masses of Single Stars}
\label{sec:problem}

\begin{figure}
\epsscale{1}
\plotone{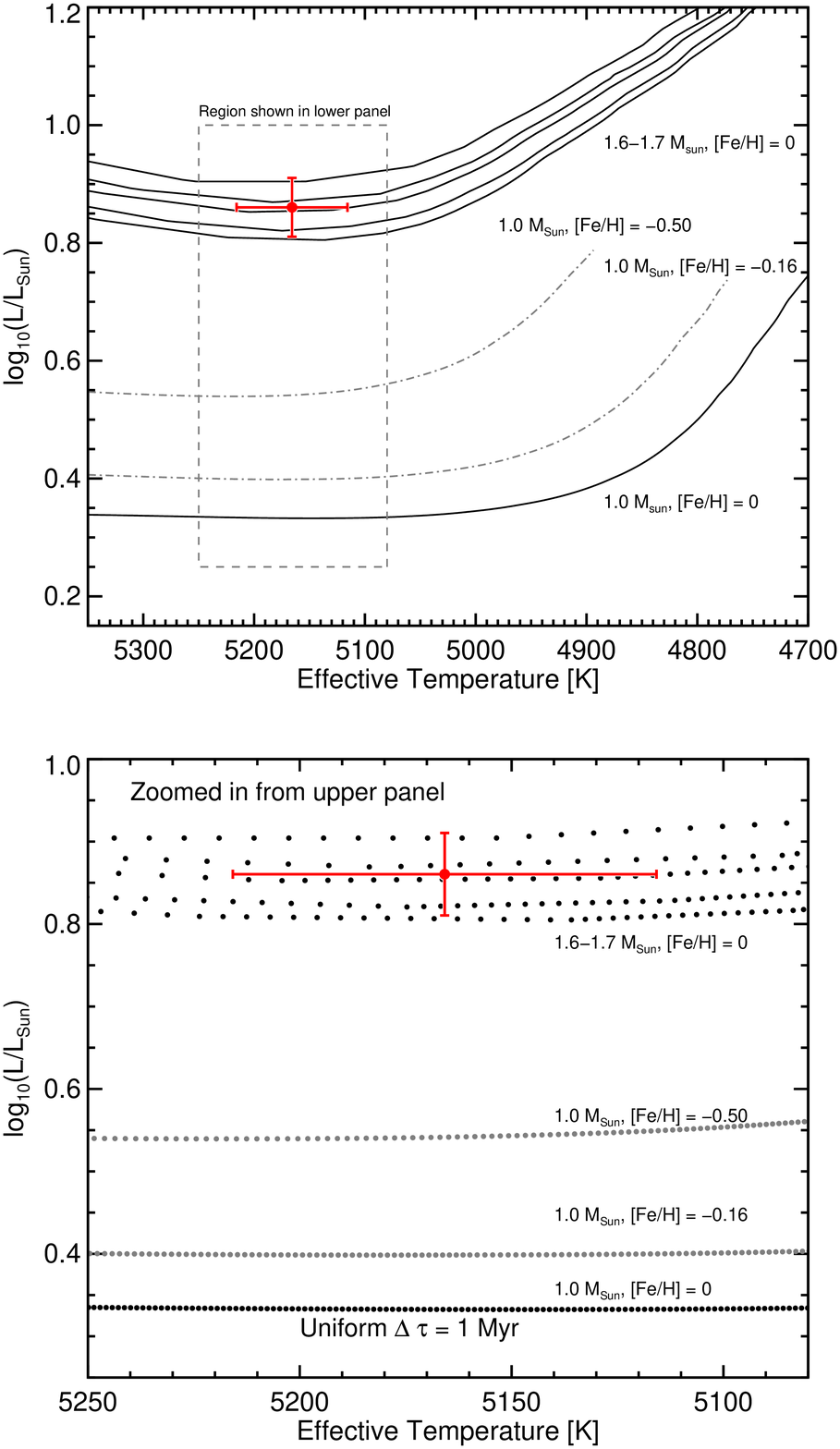}
\caption{An illustration of the interpolation of
  model-grids at the position of a subgiant's
  spectroscopically-measured \teff\ and $L$. \emph{Top:} The subgiant
  branch of the theoretical Hertzsprung-Russell diagram near the base
  of the red giant branch. The solid lines show the evolutionary paths
  of stars of various masses. The red circle shows the
  position of a specific subgiant and the error bars show the 68.2\%
  confidence region of its effective temperature and luminosity. The
  dot-dashed 
  lines show two other metallicities for the Solar-mass track,
 which we include to illustrate the effect of metallicity in this
 plane. The dashed box traces the region shown in the lower 
  panel. \emph{Bottom:}
  A zoomed-in region around the subgiant's measurements, with the mass
  tracks shown sampled at a uniform time-spacing of $\Delta \tau =
  1$~Myr, illustrating the different evolution rates for subgiants of
  various masses.
  \label{fig:tracks}}
\end{figure}

The heart of the problem is that measuring the 
masses of single stars is necessarily a model-dependent procedure. The
most common method of estimating masses is to interpolate
theoretical stellar evolution grids at the positions of various
measured stellar properties. Typically, the set of parameters used are
the stellar effective 
temperature (\teff), luminosity ($L$; or absolute magnitude
\mv) and
metallicity 
(\fe) based on LTE atmospheric models 
fitted to high-resolution stellar spectra
\citep[e.g.][]{valenti05, takeda08}. The top panel of Figure~\ref{fig:tracks}
illustrates a simplified 
interpolation in which  measurements of \teff\ and $L$ at a fixed
\fe~$=0$
(red circle and error bars) are 
compared to mass tracks from the Yale Rotational Evolution Code
models \citep[YREC;][solid black
  lines]{takeda07,demarque08}. We also demonstrate the effect of
metallicity by showing two solar-mass tracks with [Fe/H]~$ = \{-0.16,
-0.50\}$; metallicity acts as a third dimension. At a fixed mass,
metal-poor stars are hotter than 
Solar-metallicity stars, while metal-rich stars are
cooler.

\subsection{A Probabilistic Framework}

The probability of a star's mass, $M$, given its spectroscopic and
photometric 
parameters and the selection criteria of a survey is given by
Bayes' theorem

\begin{multline}
P(M | T_{\rm eff}, {\rm [Fe/H]}, L, \mathcal{M}_V, B-V, \mathcal{I})
\propto  \\
P(T_{\rm eff}, {\rm [Fe/H]}, \mathcal{M}_V~|~M)\times \\
 P(M~|~\mathcal{M}_V, B-V, \mathcal{I})  
\label{eqn:masspdf}
\end{multline}

\noindent The left-hand side of the proportionality is an expression
for the posterior probability 
distribution of the stellar mass given a spectroscopic estimate of the
stellar effective temperature \teff, metallicity \fe, and bolometric
luminosity $L$. In addition to 
the spectroscopic and photometric
properties, we also have additional information \info, which in our
analysis is
provided by the  galactic population models of \cite{girardi05}. The 
term \info\ encodes information about the stellar IMF, stellar
evolution models, and the distribution of ages and metallicities as a
function of Galactic scale height
\citep[see][for a similar application]{dawson12b}. 

The right-hand side of Eqn.~\ref{eqn:masspdf} is the product of two
probabilities. The first is the likelihood, which relates the
probability of measuring the spectroscopic properties of the star given
various choices of the stellar mass from stellar evolution models. The
second term 
describes our prior knowledge about the distribution of stellar masses
for stars throughout the Galaxy with a given range of photometric properties. 

It is common for investigators using model grid interpolations to
focus solely on the likelihood 
term, because the maximization of the likelihood is directly related
to the concept of ``chi-squared minimization'' when the measured
parameters are normally distributed. This can be seen by taking the
logarithm of the likelihood, $\mathcal{L}$, with normally-distributed
measurement uncertainties on the spectroscopic parameters in
Eqn.~\ref{eqn:masspdf}

\begin{align}
\mathcal{L} &\equiv \ln{[P(T_{\rm eff}, {\rm [Fe/H]},
    L) ]} \nonumber \\
&= C - \frac{1}{2}
(\chi^2_{\rm T_{\rm eff}} + \chi^2_{\rm [Fe/H]} +
\chi^2_L) 
\end{align}

\noindent where, e.g.

\begin{equation}
\chi^2_{\rm T_{\rm eff}} = \left[\frac{T_{\rm eff}(M) - T_{\rm eff, meas}}{\sigma_{T_{\rm eff}}} \right]^2
\end{equation}

Minimization of the $\chi^2$ terms maximizes the likelihood of the 
measurements as a function of $M$.  However, this least-squares
approach neglects prior information about the distribution of stellar
masses. 

\subsection{The Stellar Mass Prior}

\begin{figure}
\epsscale{1.0}
\plotone{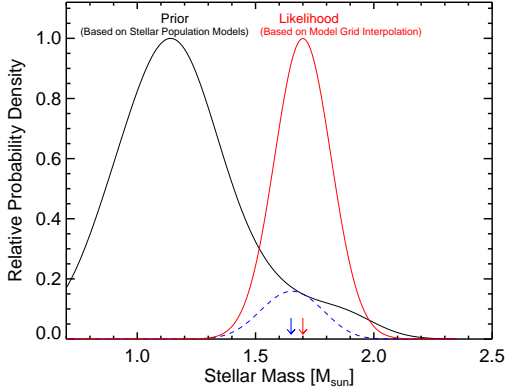}
\caption{A typical mass
  probability distribution for a 1.7~\msun\ subgiant based on
  atmospheric parameters 
  interpolated onto stellar evolution grids is 
  (likelihood; solid  red line), and the prior distribution for
  stars with comparable colors and {\it absolute} magnitudes
  (solid black line). The 
  resulting mass posterior is the product of the likelihood and prior
  (dashed blue line) and has a mean of 1.65~\msun\ (blue arrow), which
  is roughly 
  3\% lower than the initial estimate without a prior (red arrow). The
  scaling of the distributions does not matter since such factors
  scale out of the product and do not affect the centroid or symmetry
  of the final posterior distribution, hence the proportionality in
  Eqn.~\ref{eqn:masspdf}. We note that introducing an {\it apparent}
  magnitude cut would  move the mean of the prior further toward
  higher stellar masses. 
  \label{fig:probs}} 
\end{figure}

Even though the measurement
errors for stellar properties may be symmetrically distributed across
several mass tracks, 
there is not an equal likelihood of a star having masses under each
side of the measurements' probability distribution. This is both
because more massive stars are intrinsically rarer due to the IMF, and
because stars of different masses evolve at 
very different rates.  The bottom panel of Figure~\ref{fig:tracks}
illustrates this difference in evolutionary rates for the YREC models
with uniform time sampling ($\Delta \tau = 1$~Myr). Because they
evolve slower, there are many more grid points for a Solar-mass  star
on the subgiant branch than there are for more massive subgiants.  

The prior mass distribution must accurately reflect the relative
numbers of observable stars of various masses. To generate this prior
we used the  TRILEGAL Galactic stellar population simulation code that
incorporates the IMF, stellar evolution, age-metallicity relationship,
and  photometric system to produce synthetic stellar populations
\citep{girardi05}. Girardi et al. present extensive tests of
their simulated stellar populations, and demonstrate that they can
faithfully reproduce the local H--R diagram and star counts of the
\emph{Hipparcos} and 2MASS catalogs.  

We accessed the code using Perl scripts provided by L. Girardi (2012,
private communication), with the default Galactic population
parameters of the online TRILEGAL~1.5 input form. Specifically, we
assumed the \citet{chabrier03} IMF, the empirical age-metallicity
relationship measured by \citet{rocha00}, and the Padova stellar
evolution grids \citep{girardi02}. The simulation code also assumes a
multi-component Galactic stellar population, with separate
prescriptions for the thin/thick disk, bulge and halo. However, our
simulated sample of subgiants only extends to $\approx200$~pc, and
therefore only samples the immediate Solar neighborhood.

Given a particular star of 
interest, one can query the simulation to determine the 
expected distribution of masses at a given set of observed photometric
properties. In our case we use the $B-V$ color and absolute magnitude
(\mv) since these properties are available from the \emph{Hipparcos}
catalog for all subgiants in the \citet{johnson10c} target sample
\citep{hipp2}.   

In Figure \ref{fig:probs} we illustrate the effect of incorporating
such a prior into the mass measurement for a particular subgiant. We
consider a star with $B-V = 1.00 \pm 0.02$ and \mv~$= 2.2 \pm 0.5$.
The corresponding stellar mass is $M = 1.7 \pm 0.12$ \msun\ based on
spectroscopy alone, which 
represents the likelihood term.  We construct the prior using all
stars from the 
TRILEGAL simulations with similar photometric properties, using a
1-$\sigma$ cut in \mv\ and 3-$\sigma$ cut in $B-V$, to allow for enough
stars in the simulation.  We find that while this prior distribution
peaks at 1.15~\msun\, the posterior distribution resulting from the
product of the likelihood and prior peaks at 1.65 \msun.
Including the prior, which contains information about the IMF and
different evolution rates, results in a mass estimate that is 3\%
lower than the likelihood alone. Thus, the different evolution rates
of subgiants of various masses is not enough to lead to systematically
overestimate any individual subgiant's mass by $\approx50$\%, as
suggested by L11.  

Note that the prior distribution peaks near 1~\msun, similar to the
simulated mass distribution of L11. This is because the prior contains
no information about the star's actual metallicity. Once a
spectroscopic assessment of the star's metallicity (and \teff) is
made, the likelihood function modifies the prior accordingly. That,
combined with the high precision of the spectroscopic parameters
results in a likelihood term that dominates over the prior and favors
higher masses.

\section{The Mass Distribution of Subgiants in the Keck Doppler Survey}
\label{sec:survey}

We next turn our attention to the question of whether these massive
subgiants should exist at all. L11 claimed that massive, evolved
stars with $M \gtrsim 1.5$~\msun\ should be exceedingly rare.
So much so that the relative numbers of subgiants with masses in
excess of 1.5~\msun\ compared to 1.0-1.2~\msun\ should be taken as
evidence that the inferred masses must be incorrect. To test this
hypothesis, we assess the expected distribution of 
masses for stars selected in the same manner as the targets of
\citet{johnson10b}. As we will show, accounting for all selection
criteria is key to properly estimating the expected mass distribution
of sample of subgiants.

L11 simulated 
stellar populations using the YREC stellar evolution models
together with assumptions about the form of the
Galactic initial mass function (IMF). These features effectively
imposed a prior in his Monte Carlo simulations as stars
were drawn far less frequently for masses greater than 1.5~\msun\ than
those closer to Solar. In his simulations, L11 found that only 11\% of
his simulated subgiants had $M > 1.5$~\msun.

\subsection{The importance of a magnitude limit}

Stellar evolution and the IMF do not have the final say in
shaping the distribution of stellar masses for Doppler survey
targets. Surveys of subgiants 
have specific selection criteria that result in stellar samples that
are very different from the Galaxy's stellar population as a
whole. For 
example, the sample of subgiants monitored at Keck Observatory
were selected based on $0.8 < B-V < 1.05$, $1.8 < \mathcal{M}_V < 3.0$,
and $V < 8.5$ \citep{johnson10b}. Another important criterion used to
select subgiants is the 
requirement that the stars have $M > 1.3$~\msun\ when their
\emph{Hipparcos} B-V colors \citep{hipp2} and absolute V-band
magnitudes ($M_V$) are interpolated onto the Padova model
grids, under
the assumption \feh~$=0$ \citep{johnson10b}. 

A magnitude criterion was not used in the 
simulations of L11, but it has a profound impact on the expected 
mass distribution of a sample of target stars. Consider two stars,
with masses 1.2~\msun\ and 1.8~\msun. The IMF predicts a number of
stars scales as $M^{-2.4}$, and the evolution rate across the subgiant
branch scales as $M^{-3.1}$. The combined effect is a number of
subgiants that scales as $M^{-5.5}$ throughout the Galaxy. Assuming a
volume-limited survey, as L11 did, results in the expectation of an
order of magnitude more 1.2~\msun\ subgiants compared to
1.8~\msun\ subgiants. 

Now consider the volume $\mathcal{V} \sim d^3$ occupied by a star,
defined by a distance, $d \sim L^{1/2}$, out to which a star  
is brighter than the limiting magnitude of the survey. The luminosity
of stars on the subgiant branch near the base of the red giant branch
scales as $L 
\sim M^{2.5}$, based on inspection of the YREC model grids (see also
Figure 5 of L11). The volume scales with the stellar mass as
$\mathcal{V} \sim d^3 \sim L^{1.5} \sim M^4$. An apparent
magnitude cut will increase the number of \emph{observable}, massive
subgiants within a given apparent magnitude range, which partially
compensates for the dearth of  more massive stars due their shorter
evolution timescale and the stellar IMF. Imposing a magnitude
limit to the selection of stars will result a factor of
two fewer 1.8~\msun\ subgiants compared to 1.2~\msun\ subgiants. This 
is much less severe than the prediction of L11, which presumably
adopted volume-limited sample (see \S~\ref{sec:sim}).

 This effect is similar
to the Malmquist bias in galaxy surveys, in which the magnitude-limited
survey sample will result in an apparent overabundance of massive,
luminous galaxies at higher redshifts \citep{malmquist22}\footnote{As
  an historical aside, K.~G.~Malmquist also published a study of the
  distribution of (unevolved) A-type stars in the Solar
  neighborhood \citep{malmquist33}. In principle, a similar study
  could be used to check 
  the mass measurements of subgiants by comparing the ratio of A-type
  stars to the number of equally massive subgiants. This ratio should
  be equal to the ratio of the main-sequence lifetime of A dwarfs to
  the lifetime of stars on the subgiant branch.}. The simple scaling arguments
presented here give a rough sense for
the relative numbers of stars of various masses within a
magnitude-limited survey, but they do not account for all effects 
that will ultimately shape the mass distribution. For a more
thorough analysis we again turn to Galactic population models.

\subsection{Simulating the expected mass distribution of subgiants}
\label{sec:sim}

We estimate the stellar mass prior by first simulating samples of
stars over the entire sky with a wide range of apparent magnitudes.
We then select subgiants from these simulated samples in the same manner
that the retired A stars surveyed by \citet{johnson10b}, namely $0.8 <
B-V < 1.05$, $1.8 <$~\mv~$<3$, $V < 8.5$ and the restriction that the
stars' colors and absolute magnitudes correspond to $M >
1.3$~\msun\ based on Solar-metallicity stellar models.

We simulated 768 lines of sight, uniformly distributed across the sky
using the Hierarchical Equal Area isoLatitude Pixelization (HEALPIX) 
scheme\footnote{\url{http://healpix.jpl.nasa.gov/}}, with the
extinction at infinity calculated by the NASA/IPAC extragalactic
database \citep{schlafly11}. We avoided the galactic plane ($|b| <
4^\circ$) because TRILEGAL is known to exhibit discrepancies
with observational surveys in the Galactic plane \citep{girardi05},
and because of the large number of stars returned by those
simulations. 

\begin{figure}
\epsscale{1}
\plotone{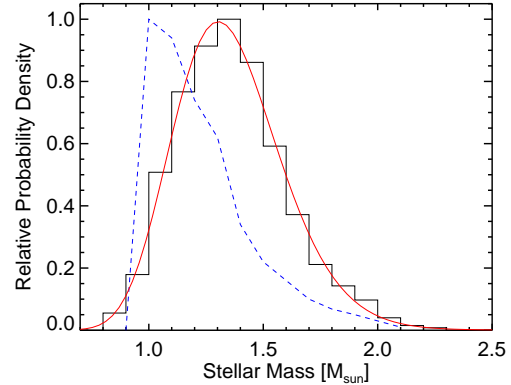}
\caption{Simulated sample of stars selected with the same criteria as
  the subgiants in RV surveys (black histogram). The best-fitting
  log-normal distribution is also shown (red). For comparison, we have
  also shown the mass distribution reproduced from Figure 4 of L11
  (blue dashed lines), which is based on a volume-limited survey
  rather than a magnitude-limited survey such as ours.
  \label{fig:massdist}}
\end{figure}

Figure~\ref{fig:massdist} shows the resulting distribution 
of simulated subgiant masses. For ease of future use, we adopt an
analytic form of the distribution; we find the posterior distribution
is described well by the expression 

\begin{equation}
P(M~|~\mathcal{I}) \approx \frac{1 + sz}{\sqrt{2\pi \sigma_M^2}} \exp{\left(-\frac{z^2}{2}\right)}
\end{equation}

\noindent where

\begin{equation}
z = \frac{\log_{10}(M) - M_0}{\sigma_M}
\label{eqn:prior}
\end{equation}

\noindent and $M_0 = 0.134$~\msun, $\sigma_M = 0.0783$~\msun\ and the
dimensional skew term is $s = -0.253$, for all masses $M >
0.7$~\msun. 

Our stellar mass distribution is qualitatively similar to that shown
in Figure 4 of L11. However, while the L11 distribution peaks sharply at
1.0~\msun, ours has a peak near 1.3~\msun, with relatively few
Solar-mass subgiants. Indeed, there are twice as many stars in our
simulations with
$M > 1.5$~\msun\ as there are for $M < 1.1$~\msun. The
reason for the differences between our simulation and 
L11's is the apparent 
magnitude cut, and the a priori selection of stars that reside near
model tracks corresponding to $M > 1.3$~\msun. While the IMF and
subgiant evolution rate 
favor less massive stars, the higher 
luminosities of more massive subgiants makes them visible within a
much larger volume. 

To test the effects of our added selection criteria compared to L11,
we selected stars by relaxing certain cuts. By relaxing the 
criterion $M > 1.3$~\msun\ when stars' \mv\ and $B-V$ colors are
compared to Solar-metallicity model grids, we  
find that the low-mass tail of the distribution is filled in, which
brings the peak of the posterior distribution down to 1.2~\msun. 
When we impose a volume limit, we recover a distribution
very similar to the volume-limited sample shown in 
Figure~4 of L11.

\section{Conclusions}
\label{sec:results}

L11 argued that the mass measurements of subgiants with $M >
1.5$~\msun, i.e. the retired A stars surveyed by \citet{johnson10b},
must be in error because stars in this mass range should be exceedingly
rare. L11 further argues
that stellar 
evolutionary models are sufficiently ambiguous in their predictions
(given reasonable uncertainties in their input physics) and that
spectroscopically determined stellar parameters are subject to large
systematic errors. L11 concludes that the true masses of the 
stars in the Johnson et al. sample are more reasonably estimated to be
1.0-1.2 solar masses, not typically closer to 1.5 solar masses. 

We applied the survey selection criteria of Johnson et al. to the
TRILEGAL galactic synthesis models and have shown that the resulting
simulated target sample has a mass
distribution consistent with the Johnson et al. mass measurements and
inconsistent with the prediction of L11.  L11 may be correct that
stellar evolution and Galactic synthesis models have substantial
uncertainties. However, since the TRILEGAL models successfully
and accurately reproduces the stellar characteristics of stars in the
Solar Neighborhood \citep{girardi05}, we find no reason to doubt their
accuracy at the level that would implicate the Johnson et al. mass
measurements. 

Nevertheless, tests of systematic errors in stellar evolution models
using planet transit light curves, eclipsing binaries and
asteroseismology are very much worthwhile. Fortunately, the large
number of transiting planets  and eclipsing binaries in the NASA
\kep\ mission target field \citep{prsa11},  together with
the exquisite photometric precision produced by the \kep\ space
telescope, will provide many opportunities for these tests in the
near future.  

\acknowledgments We gratefully acknowledge the insightful comments and
edits of earlier drafts of this manuscript provided by Rebekah
Dawson, Kaitlin Kratter, Geoff Marcy, Ed Turner and Jon Swift. We
thank James Lloyd for 
his comments and assistance with understanding the arguments of
L11, and also for his many collegial discussions and thoughtful
feedback on previous drafts of this manuscript. We also thank Leo
Girardi for  sending us his Perl scripts and    
providing advice and information about the TRILEGAL code. J.A.J. 
acknowledges support from the Alfred P. Sloan Foundation, and the David
and Lucile Packard Foundation. The Center for Exoplanets and Habitable
Worlds is supported by the 
Pennsylvania State University, the Eberly College of Science, and the
Pennsylvania Space Grant Consortium.

\end{document}